# Experimental observation of highly anisotropic elastic properties of two-dimensional black arsenic

Jingjing Zhang,[1,2,#] Shang Chen,[1,#] Guoshuai Du,[1,2,#] Yunfei Yu,[1,2] Wuxiao Han,[1,2] Qinglin Xia,[3] Ke Jin,[1,*] and Yabin Chen[1,2,4,*]

[1]*Advanced Research Institute of Multidisciplinary Sciences (ARIMS), Beijing Institute of Technology, Beijing 100081, P.R. China*

[2]*School of Aerospace Engineering, Beijing Institute of Technology, Beijing 100081, P.R. China*

[3]*School of Physics and State Key Laboratory of Powder Metallurgy, Central South University, Changsha, Hunan 410083, P.R. China*

[4]*BIT Chongqing Institute of Microelectronics and Microsystems, Chongqing, 400030, P. R. China*

[*]Correspondence and requests for materials should be addressed to: chyb0422@bit.edu.cn (Y.C.), Jinke@bit.edu.cn (K.J.)

[#]These authors contributed equally to this work.

**ABSTRACT:** Anisotropic two-dimensional layered materials with low-symmetric lattices have attracted increasing attention due to their unique orientation-dependent mechanical properties. Black arsenic (b-As), with the puckered structure, exhibits extreme in-plane anisotropy in optical, electrical and thermal properties. However, experimental research on mechanical properties of b-As is very rare, although theoretical calculations predicted the exotic elastic properties of b-As, such as anisotropic Young's modulus and negative Poisson's ratio. Herein, experimental observations on highly anisotropic elastic properties of b-As were demonstrated using our developed *in situ* tensile straining setup based on the effective microelectromechanical system. The cyclic and repeatable load-displacement curves proved that Young's modulus along zigzag direction was ~1.6 times greater than that along armchair direction, while the anisotropic ratio of ultimate strain reached ~2.5, attributed to hinge structure in armchair direction. This study could provide significant insights to design novel anisotropic materials and explore their potential applications in nanomechanics and nanodevices.

**INTRODUCTION**

Two-dimensional (2D) layered materials have aroused extensive research interests and displayed plentiful of applications in functional nanodevices and composite materials, due to their distinguished electronic, optical and mechanical properties.[1-5] Considering the unique planar structure without any dangling bonds on the surface, 2D layered materials generally present unique mechanical behaviors, such as super flexibility, ultrahigh Young's modulus, and huge elastic strain limit.[6,7] It is reported that thousands of 2D materials are proposed by high-throughput calculations,[8] however, most of which possess high symmetric structures and thus show the in-plane isotropic properties, that is, orientation-dependence is rather weak or negligible.[9] For instance, Young's modulus of graphene (point group $D_{6h}$) is almost identical along zigzag (ZZ) and armchair (AC) directions, as demonstrated in stress-strain relationship by uniaxial tensile tests.[10] In contrast, anisotropy as a novel degree of freedom, enables us to explore low symmetric 2D materials, and thus their anisotropic mechanical properties.[11] The representative anisotropic 2D materials, such as black phosphorus (b-P) and SnSe, exhibit the fantastic negative Poisson's ratio, and distinguished elastic modulus along AC and ZZ directions.[12,13] To the best of our knowledge, most research in this aspect are limited to theoretical calculations.[14]

Unlike bulk materials, great challenges arise to experimentally measure the mechanical properties of 2D materials owing to the atomic thickness and complex fabrication procedures.[15] Generally, nanoindentation method based on atomic force microscopy (AFM) is widely utilized to investigate the elastic properties of graphene, $MoS_2$, and $WSe_2$.[16-18] In this method, the tunable force applied by a sharp tip on cantilever, allows us to deform the 2D flakes suspended on a trench or hole. Young's modulus and pretention can be derived based on the proposed circular drum model or doubly clamped beam model.[19-21] However, the testing load is rather concentrated around the indenter tip and the 2D flakes normally collapse, which seriously affects the uniformity of strain and thus the accuracy. In addition, the two-point or four-point bending tests based on a flexible polymer substrate surfer from the indirect determination of strain and stress, which benefits strain-regulated optical and electrical measurements.[22,23] Therefore, the *in situ* microelectromechanical system (MEMS)-based tensile testing methods are developed to achieve uniform in-plane stress, and real-time recording of load-displacement data and structural morphology. In this regard, it was found that monolayer graphene can withstand the elastic strain up to ~6%,[24] and hexagonal boron nitride (*h*-BN) presented the defect-tolerated and robust mechanical properties.[25]

Black arsenic (b-As, space group *cmca*), as a novel metastable semiconductor, has emerged in recent years and displayed extreme in-plane anisotropic electrical, optical and thermal properties due to its puckered structures.[26-28] Importantly, it is reported that b-As showed relatively better ambient stability than b-P, asymmetric Rashba valley and exotic quantum Hall effects.[29-31] The theoretical calculations predicted that b-As exhibited highly anisotropic mechanical properties, including elastic modulus, tensile stiffness, and ultimate strain. For example, Young's modulus along ZZ direction is almost three times higher than that along AC direction, leading to the diversely stiff and flexible nature along ZZ and AC direction, respectively.[32] Notably, b-As also showed the negative Poisson's ratio as −0.09,[33] and the auxetic effect provides its wide applications in shear resistance and energy absorption. Despite much effort, there is still lack of experimental evidence on the anisotropic mechanics of 2D b-As.

In this work, we performed *in situ* elastic straining tests of 2D b-As based on MEMS devices under scanning electron microscopy (SEM), and the cyclic load-displacement curves demonstrated the highly anisotropic elastic properties of 2D b-As. The suspended b-As specimen with specific orientation was identified by polarized Raman spectroscopy, and followed by focused ion beam (FIB) technique to pattern into nanoribbon. It is found the Young's modulus of ~40.8 GPa in ZZ direction was ~1.6 times higher than that along AC direction (~25.8 GPa), and the anisotropic ratio of ultimate strain was up to ~2.5. We believe that 2D b-As with highly anisotropic properties could be an insightful cornerstone for fundamental research of anisotropic 2D materials and their broadly promising applications.

**RESULTS AND DISCUSSION**

**Preparation and characterization of suspended b-As nanoribbon**

Preparation of such a suspended b-As nanoribbon sample for *in situ* elastic straining measurement is with great challenge. B-As material from the natural mineral was utilized, and the surface appeared shiny luster. To study the anisotropic elastic properties of 2D b-As, its ZZ or AC orientation needed to be precisely aligned with the tensile direction, and a top-down fabrication process was developed, as shown in Figure 1a. First, we prepared the appropriate b-As flake on Si wafer using polydimethylsiloxane (PDMS)-assisted exfoliation method, which facilitated the large area, clean surface and uniform thickness of the obtained flake. Lattice orientation of b-As flake was successfully identified by the mature angle-resolved polarized Raman spectroscopy, owing to the distinct selection rules and Raman tensors of phonon modes (the results shown in Supplementary Figure S1). Second, FIB technique with the prominent spatial resolution was used to pattern a given b-As flake to rectangular nanoribbons along ZZ and AC directions, and the desired length and width were around 10 and 1 μm, respectively. The typical images can be found in Supplementary Figure S1. Afterwards, the fabricated b-As nanoribbon was gently transferred to bridge two MEMS arms using a sharp probe tip aided with micro-manipulator. It is noted that this dry-transfer method benefits the fine orientation adjustment of b-As nanoribbon perpendicular to MEMS arm, in order to exclude the potential shearing effect. In final, two ends of the aligned b-As nanoribbon were bonded to MEMS arms by the metallic Pt using FIB, in order to prevent any slippage of b-As from arm surface. The deposition condition was optimized to minimize the possible damage of b-As specimen. Moreover, most fabrication steps were carried out in vacuum, and the entire process was with very limited period of exposure in air to warrant the minimum oxidization or degradation of b-As crystal.

MEMS-based nanoindentation system[34] was employed to study the anisotropic elastic properties of 2D b-As. Figure 1b shows a representative SEM image of the well-fabricated b-As sample. It can be seen that the tensile load was actually applied to b-As nanoribbon by stretching the movable body through a micro-hook above, which were suspended by four flexures within an outer fixed frame. This micro-hook can be directionally connected with the other micro-hook mounted on the nanomechanical testing system (more details in Supplementary Figure S2). This MEMS-based nanoindentation technology can provide a tunable force up to 200 mN and the unique resolution of displacement as high as 0.05 nm. Obviously, surface of the central b-As nanoribbon remained free of any external contaminations or structural damages as shown in Figure 1c, allowing us to investigate the intrinsic mechanical properties of the natural b-As.

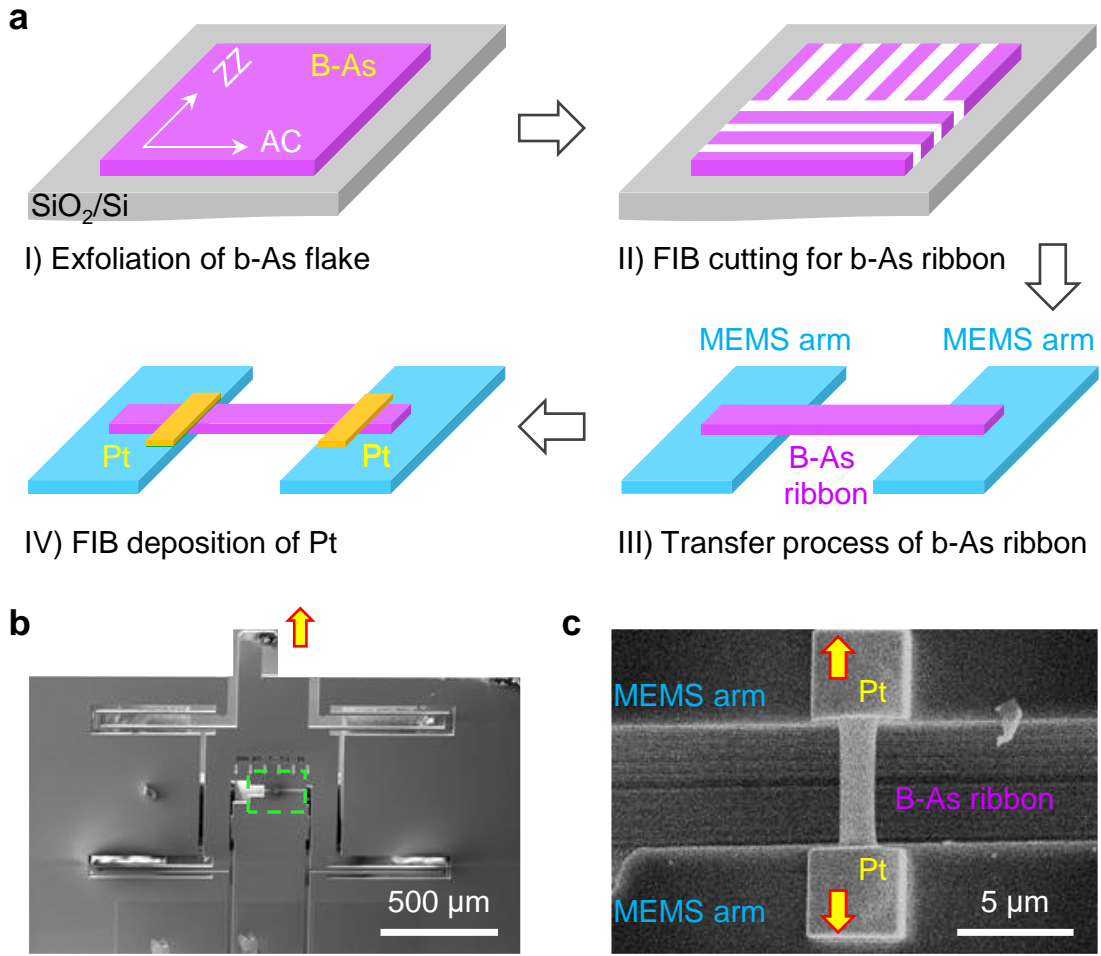

**Figure 1.** Sample preparation method for *in situ* elastic straining test of the suspended b-As flake under SEM. **a)** Schematic illustration of the whole fabrication process. In general, four steps are as follows: I) mechanical exfoliation of the clean, uniform and large b-As flake on Si wafer. The orthogonal AC and ZZ orientations can be identified by Raman spectroscopy. II) FIB technique was used to gently shape the large b-As flake into nanoribbons (purple parts) along AC and ZZ directions, respectively. The white areas were removed away, which resulted in the straight and sharp edges of b-As flake. III) dry transfer of the prepared b-As nanoribbon onto the MEMS device to bridge two suspended arms. The transfer process was realized by a sharp tip. IV) FIB deposition of Pt (yellow strips) on the b-As to enhance its interaction with MEMS arms, in terms of the potential slippage. **b)** The fabricated b-As samples for *in situ* tensile test with the detailed structures of MEMS device. To stretch the upper hook indicated by yellow arrow, tensile load can be continuously applied to the central b-As flake. **c)** Zoom-in view of the interior structure as marked in the green area in **b)**. Two yellow arrows point to their stretching directions.

## *In situ* tensile test of 2D b-As along different directions

To study its orientation-dependent elastic properties, we successively demonstrated *in situ* tensile tests on the individual b-As nanoribbon along AC and ZZ directions. The displacement rate as low as 2.5 nm/s offered a quasi-static state of MEMS device and b-As crystal. As displayed in Figure 2a to 2d, the representative SEM images captured in the recorded video showed the elastic deformation of b-As nanoribbon along AC direction. Notably, the whole surface of b-As sample remained very clean during *in situ* tensile test. As the tensile load continuously increased, the engineering elastic strain gradually grew from 0% to 2.7%, 4.4%, and 6.2%, without any new defect or crack. After unloading, the b-As nanoribbon totally returned to its initial position, indicating that AC direction of b-As presented the reasonable elastic response. Moreover, the different load-displacement curves exhibited very good linear behavior till the maximum force reached 720 µN in this measurement, as seen in Figure 2e. Importantly, it is obvious that all curves were well consistent with each other, suggesting the high stretchability nature of b-As crystal and the mature fabrication process of testing device as well.

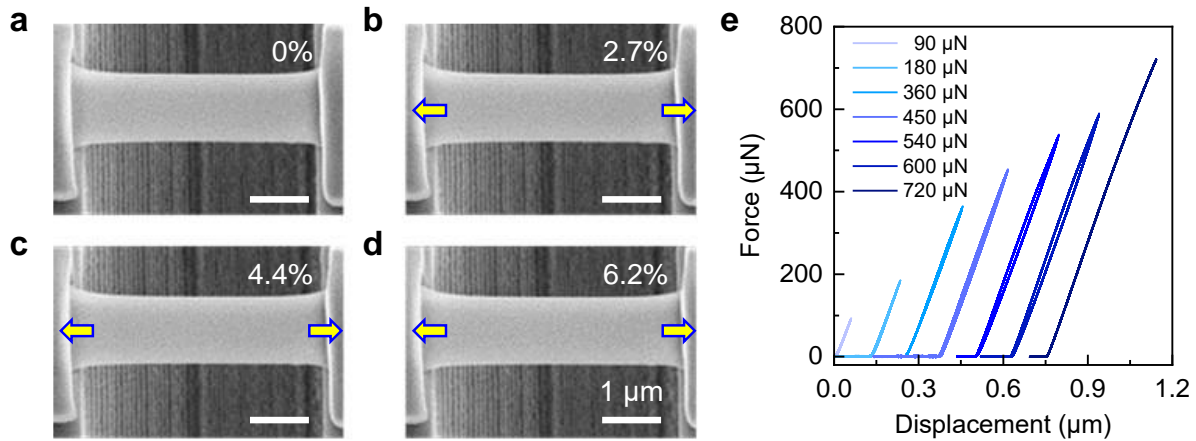

**Figure 2.** *In situ* tensile test and elastic straining of b-As nanoribbon along AC direction. **a-d**) SEM images captured in recorded video during tensile test. The thickness, gauge length and width of this suspended b-As nanoribbon are 239.4 nm, 5.1 µm and 1.5 µm, respectively. The calculated engineering elastic strains ranged from 0% (**a**) to 2.7% (**b**), 4.4% (**c**), and 6.2% (**d**). The load was continuously applied horizontally, as indicated by the yellow areas in each image. **e**) The load-displacement curves of b-As nanoribbon with tensile strains along AC direction. The forces were tuned from 90 to 720 µN. All curves are well consistent with each other, suggesting the high stretchability nature of b-As crystal. The curves were shifted horizontally for clarity.

Next, we performed the *in situ* tensile test of b-As along ZZ direction, and the b-As nanoribbon was fabricated from the same flake as above. As can be seen in Figure 3a to 3d, with the increase of tensile load along ZZ direction, the displacement (*i.e.* tensile strain) of b-As nanoribbon gradually became larger, and the corresponding engineering elastic strain rose from the initial 0% to 0.8%, 1.2%, and 1.8%, respectively. During the whole loading process, no apparent cracks or defects appeared, and thus b-As nanoribbon remained the crystalline structure. As displayed in Figure 3e, the corresponding force increased linearly with the recorded displacement in each test, and all curves showed great consistency, suggesting that 2D b-As has high elastic repeatability. The hysteresis in each loading-unloading curve maybe contributed from mechanical energy dissipation or the external indenter of MEMS device.[35] The maximum force reached 500 μN in this case, while the force-displacement slope appeared larger than that along AC direction, indicating the greater tensile stiffness and elastic modulus in ZZ direction.

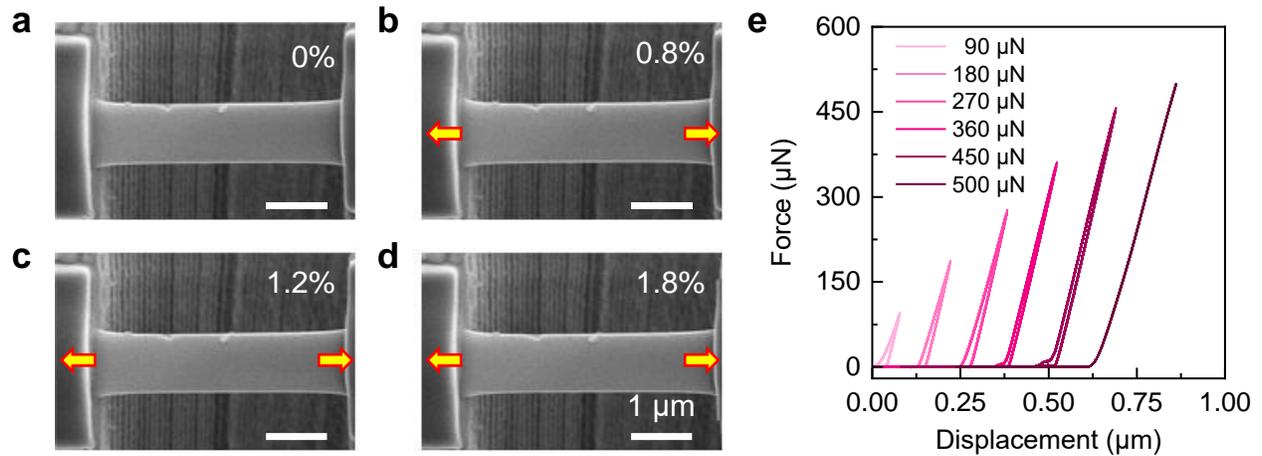

**Figure 3.** Elastic deformation of b-As nanoribbon in ZZ direction. **a-d**) SEM morphologies of the b-As specimen with 0%, 0.8%, 1.2% and 1.8% deformation, respectively. The b-As nanoribbon was stretched horizontally, as indicated by the yellow arrows. The initial thickness, gauge length and width of this suspended b-As nanoribbon are 239.4 nm, 5.3 μm and 1.2 μm, respectively. **e**) Load-displacement curves of b-As nanoribbon under different strains in ZZ direction. The applied force varied from 90 to 500 μN. All curves present linear behavior and well consistent with each other. The curves were shifted horizontally for clarity.

Thereafter, we turned to examine the ultimate stretchability and fracture behavior of b-As nanoribbon along different directions. The indenter force has been continued to increase until the sample cracked completely. In Figure 4a, the load-displacement curves displayed three distinct stages. In the first stage, each curve seemed flat with a relatively small or ignored slope, corresponding to the micro-hooks moving from their separation to the right connection, which further indicated the b-As nanoribbon kept straight at the initial position without apparent collapse, quite different from the pre-relaxed graphene or h-BN monolayers.[24,25] In the second stage, b-As nanoribbon started to be actually stretched as the displacement increased, while the indenter load acted on both b-As sample and MEMS device simultaneously. Along AC direction in Figure 4a, we can extract the slop $k = dF/dd$ (F and d represent the load and displacement, respectively) as ~1842.9 N/m from the linear curve, attributed to the sum of tensile stiffness from b-As and MEMS device. Then, the load dropped abruptly when exceeding 815 μN, suggesting that brittle fracture of b-As occurred, as marked by the red asterisk. In the third stage, the load rose slowly and only acted to drive MEMS device, and its tensile stiffness was linearly fitted as ~24.3 N/m. In this way, the intrinsic tensile stiffness of b-As along AC direction can be extract as ~1818.6 N/m, by subtracting the contribution from MEMS device. It is reasonable to assume that b-As nanoribbon was under uniaxial stress condition, the three-dimensional (3D) Young's modulus of tensile specimen can be expressed as $E_{3D} = kl/tw$ in solid mechanics, where $k$, $l$, $t$, and $w$ represents the tensile stiffness, bridging length, thickness, and width of b-As nanoribbon, respectively. In final, we obtained the $E_{3D}$ as ~25.8 GPa for b-As along AC direction. Similarly, in the case of ZZ direction, the fracture behavior of b-As nanoribbon happened when the maximum load exceeded 525 μN. The fitted tensile stiffness for the entire system and MEMS device were 2269.5 and 55.1 N/m, respectively, leading to the intrinsic stiffness of b-As as 2214.4 N/m. In final, the $E_{3D}$ along ZZ direction was calculated as ~40.8 GPa, and the anisotropic ratio reached ~1.6, indicating the highly anisotropic elastic properties of b-As crystal.

According to the uniaxial stress hypothesis in continuum mechanics,[24] the tensile stress $\sigma$ was related with load F with $\sigma = F/A$, where A represented the cross-sectional area of b-As sample. In principle, the tensile strain $\varepsilon$ can be acquired by $\varepsilon = \Delta l/l$, where $\Delta l$ means the elongation of the b-As nanoribbon. Thus, engineering stress-strain curves of b-As in different directions can be obtained, as shown in Figure 4b. The extracted Young's modulus of ~44.1 and ~27.2 GPa along

ZZ and AC directions are quantitatively consistent with that in Figure 4a, respectively. In addition, we can examine the catastrophic fracture phenomena of b-As as shown in Figure 4c and 4d. The brittle fracture behaviors along both AC and ZZ orientations mostly happened perpendicularly to their tensile direction, leaving two separated pieces with sharp propagation paths. Importantly, the elastic strain limit along AC direction around 8.9% was ~2.5 times larger than that along ZZ direction. Also, the ultimate stretchability of b-As is evidently better than graphene (~6%)[24] and h-BN (~6.2%) monolayers,[25] owing to its puckered lattice structure. We can see that facture behavior occurred near the clamped edge in Figure 4c, which possibly originates from the stress concentration. This edge effect maybe weakened by optimizing FIB condition.

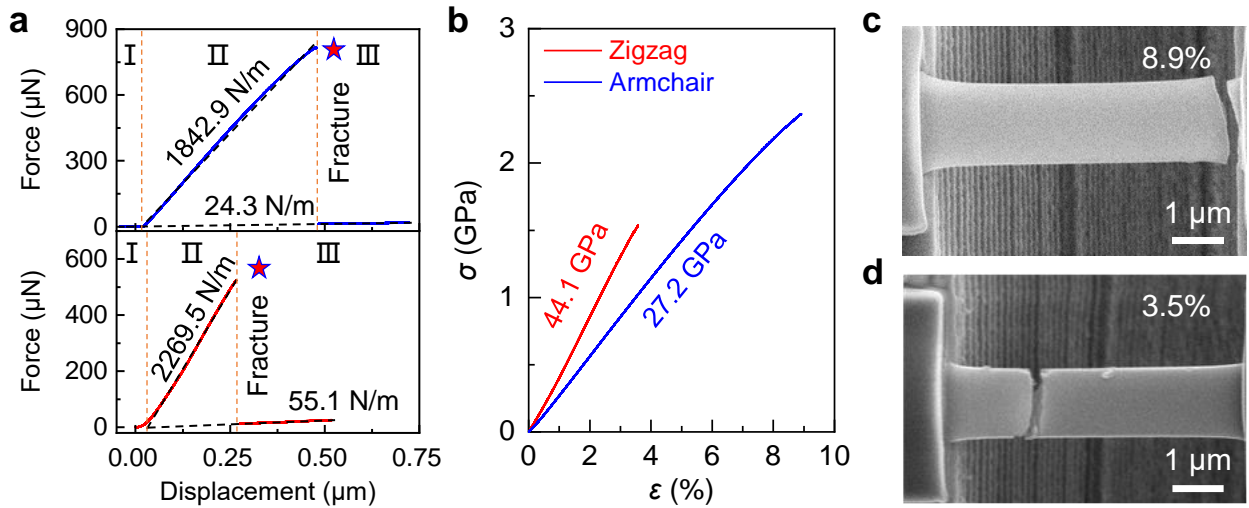

**Figure 4.** Ultimate stretchability and fracture behavior of b-As nanoribbon along different directions. **a**) Load-displacement curves of b-As nanoribbon in the AC (upper) and ZZ (lower) directions. There are three distinct stages (I, II and III) in each curve, separated by vertical dished lines. The red asterisks indicate where the brittle fracture occurred. The tensile stiffness of b-As and MEMS device were obtained by linear fitting (black dished lines). **b**) The extracted engineering strain–stress ($\sigma$–$\varepsilon$) relationships of b-As along ZZ (red) and AC (blue) directions. The fitted Young's modulus $E_{3D}$ is ~44.1 GPa (~27.2 GPa) along ZZ (AC) direction. **c**) The representative SEM image of the fracture morphology of b-As nanoribbon along AC direction, and the ultimate strain was up to ~8.9%. **d**) SEM result after catastrophic brittle fracture of b-As along ZZ direction. The ultimate strain reached ~3.5%.

**Highly anisotropic elastic properties of 2D b-As**

Furthermore, it is noticed that 2D b-As presented the highly anisotropic elastic properties and fracture behavior. The schematic diagram of its puckered structure is depicted in Figure 5a. Each As atom is bonded with three neighbors by two distinct covalent bonds, the longer in-plane one and the other shorter out-of-plane one. Two parallel atomic layers of b-As eventually contribute to the puckered honeycomb structure with the orthogonal AC and ZZ directions. In term of nanomechanics, mechanical properties of b-As are reasonably expected to display the high anisotropy owing to its puckered lattice and low symmetry. The smaller lattice constant along ZZ direction suggests its stronger modulus and fracture strength. In contrast, b-As along armchair direction can display the superior tensile resistance and thus withstand the larger strain limit.

As show in Figure 5b, we summarized the elastic properties of many typical 2D nanomaterials,[32,36,37] including b-P,[38-40] transition metal dichalcogenides (MX$_2$, M=Mo or W; X=S, Se, or Te),[41] graphene,[10,42] C-B-N compounds,[43-45] and IV-V semiconductors.[46] To the best of our knowledge, the experimental work was very limited. The anisotropy of elastic properties can be evaluated via the ratios of Young's modulus ($E_{3D}^{AC}/E_{3D}^{ZZ}$) and ultimate strain ($\varepsilon_c^{AC}/\varepsilon_c^{ZZ}$) along AC and ZZ directions, respectively. Notably, graphene and $h$-BN kept the maximum elastic modulus as ~1TPa, while these highly symmetric materials present isotropic mechanical properties. Even though the experimental Young's modulus of b-As are reasonably weaker than that of b-P, due to its smaller As-As bond strength together with larger lattice cell, the measured anisotropic $E_{3D}^{AC}/E_{3D}^{ZZ}$ was obviously higher than few layer b-P,[39] which is well agreement with theoretical results.[47] To be honest, we were unable to measure the mechanical properties of few-layer or monolayer b-As, and even further to discuss its potential size effect on Young's modulus and fracture behavior, mostly due to the great challenge of device fabrications. More advanced fabrication approach is essential for this issue. Alternatively, machine-learning interatomic potentials emerge these years as an exotic approach,[48,49] and could be utilized to calculate mechanical properties of b-As.

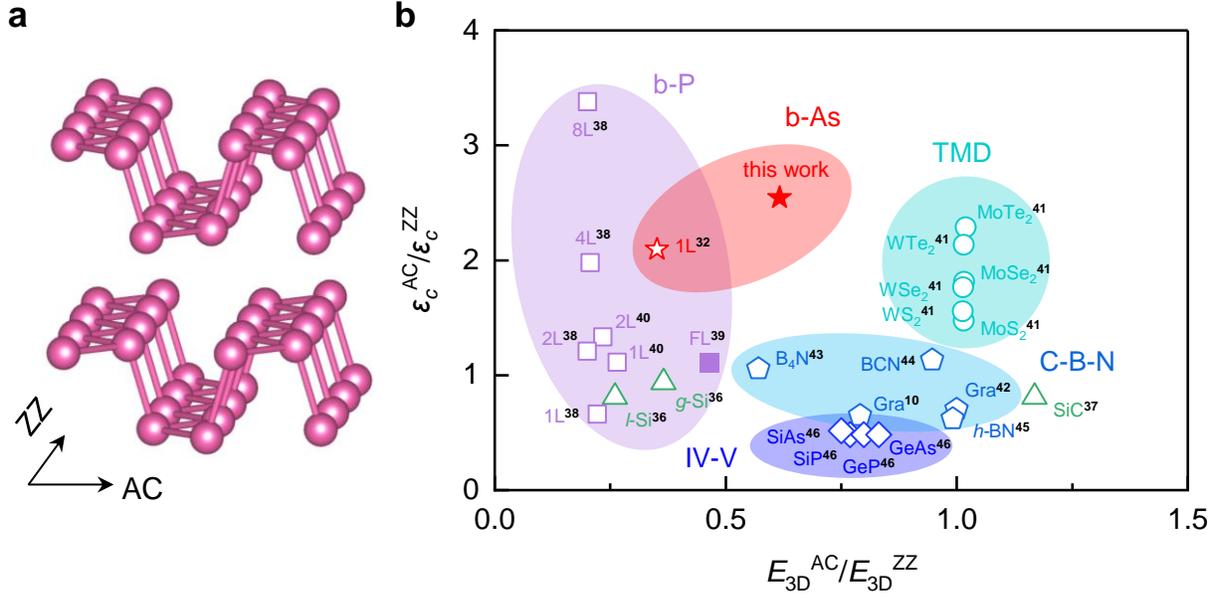

**Figure 5.** Highly anisotropic elastic properties of 2D b-As, compared with other typical 2D materials. **a**) The atomic lattice of b-As with puckered structure. **b**) Summary of anisotropic Young's modulus and strain limit of various 2D materials. The filled and empty symbols indicate the results from experiments and theoretical calculations, respectively. The calculated elastic properties of b-P showed layer-number dependence, and the few-layer (FL) result were measured by AFM indentation method. All references are labeled with black numbers.

**CONCLUSION**

In summary, anisotropic mechanical properties of 2D b-As have been investigated using the established *in situ* tensile testing setup under SEM, which included the flake exfoliation, FIB cutting, dry-transfer method and real-time recording of load-displacement curves. It was found that 2D b-As exhibited high anisotropic elastic properties along its AC and ZZ directions, owing to the puckered lattice structure. The measured Young's modulus along ZZ and AC directions was around 40.8 and 25.8 GPa, respectively, while the AC direction can significantly withstand about 2.5 times larger ultimate strain. In comparison with other typical 2D materials, the remarkable mechanical anisotropy of b-As could shed a light on the fundamental reaseaches together with many promising applications of anisotropic 2D materials.

**EXPERIMENTAL METHODS**

**Sample exfoliation and characterization of b-As**

The b-As flakes for tensile tests were meachnically exfoliated onto silicon wafer surafce (300 nm oxidazation layer) by using the PDMS stamp asstisted method. The flexible PDMS template was free of any potential residue of adhexive tape, which inevitablely happens in Scotch tape. The lattice orientaions of b-As flake were identified with angle-resolved polarzied Raman spectrum, further confirmed by transimmsion electron microscope with atomic resolution as reported elsewhere.[26,50] All Raman characterizations were conducted using Horiba iHR550 setup, and the great spectral resolution was better than 0.1 cm$^{-1}$, offered by 1800 gmm$^{-1}$ grating. The wavelength of excitation laser was 532 nm, and the diameter of focused beam approached 1 μm with 100X objective. The laser power was as weak as ~20 μW to prevent the possible chemical degradation induced by thermal effect.

**FIB cutting and Pt deposition**

The exfoliated b-As samples were cut into rectangular strips along two lattice orientations using FIB on the FEI Helios G4 UC dual beam system. The operating voltage and current of the focused beam were 5 kV and 0.8 nA, respectively. The b-As nanoribbons were then carefully transferred onto the stretch-chip channels (5 μm in width) with the aid of a probe under an optical microscope. Two ends of b-As nanoribbon were clamped on channel arms of the stretch-chip using Pt deposition method (10 kV, 7.7 pA) inside the FIB.

***In situ* streching test under SEM**

Uniaxial tensile tests were performed at room temperature with a displacement rate of 0.025 μm/s in the displacement control mode (Find Contact to realize the connection between hooks) using a nanomechanical testing system (model FT-NMT04, FemtoTools, Buchs, Switzerland). The force was measured with a MEMS-based micro-force sensor (model FT-S20000), and *in situ* imaging was performed in SEM (FEI Helios G4 UC dual beam system) to observe the tensile deformation and fracture process.

**Data Availability**

All data that support the findings of this study are available from the corresponding author upon reasonable request.

**Acknowledgements**

This work was financially supported by the National Natural Science Foundation of China (grant numbers 52072032, 12090031, and 12274467), the 173 JCJQ program (grant number 2021-JCJQ-JJ-0159), and the National MCF Energy R&D Program (2022YFE03120000). Prof. Q.L. Xia acknowledges the funding support by State Key Laboratory of Powder Metallurgy in Central South University.

**Author Contributions**

Y.C. conceived this research project and designed the experiment. J.Z., G.D., Y.Y., W.H., and Q.X. prepared the flake samples and further identified their lattice orientations. S.C. and J.Z. carrier out the *in situ* tensile tests advised by K.J. Y.C. G.D., and J.Z. wrote the manuscript with the essential input of other authors. All authors have given approval of the final manuscript.

**Competing interests**

The authors declare no competing financial interests.

**Supplementary Information**

The Supplementary Information is available and free of charge online.

Optical images of the exfoliated b-As flake, the fabricated nanoribbon, and polarized Raman spectrum. The details of *in situ* mechanical testing system with micro-force sensing probe, MEMS device and micro-hooks. More experimental data about the elastic straining tests of b-As nanoribbons. The schematic illustration of the rupture edges of b-As nanoribbons.